\documentclass[epj]{webofc}
\usepackage[utf8]{inputenc}
\usepackage[varg]{txfonts}   
\usepackage{booktabs}
\usepackage{xcolor}
\usepackage{float}
\definecolor{darkred}{rgb}{0.4,0.0,0.0}
\definecolor{darkgreen}{rgb}{0.0,0.4,0.0}
\definecolor{darkblue}{rgb}{0.0,0.0,0.4}
\usepackage[bookmarks,linktocpage,colorlinks,
    linkcolor = darkred,
    urlcolor  = darkblue,
    citecolor = darkgreen]{hyperref}
\usepackage{subcaption}
\wocname{EPJ Web of Conferences}
\woctitle{Lattice2017}
\begin{document}
%
\selectlanguage{english}
\title{%
Dual simulation of the massless lattice Schwinger model with topological term and non-zero chemical potential
}
\author{%
\firstname{Daniel}  \lastname{G\"oschl}\inst{1}\fnsep\thanks{Speaker, \email{daniel.goeschl@uni-graz.at}}
}
\institute{%
Institute for Physics, University of Graz, Austria 
}
\abstract{%
We discuss simulation strategies for the massless lattice Schwinger model with a topological term and finite chemical potential. The 
simulation is done in a dual representation where the complex action problem is solved and the partition function is a 
sum over fermion loops, fermion dimers and plaquette-occupation numbers. We explore strategies to update the fermion loops
coupled to the gauge degrees of freedom and check our results with conventional simulations (without topological term and at 
zero chemical potential), as well as with exact summation on small volumes. Some physical implications of the results are discussed.
}
\maketitle
\section{Introduction}\label{intro}

In recent years dual representations have been successfully used to overcome 
complex action problems for a variety of lattice field theories (see, e.g., the reviews 
\cite{Chandrasekharan:2008gp,deForcrand:2010ys,Wolff:2010zu,Gattringer:2014nxa,Gattringer:2016kco}). 
For further developing these techniques the Schwinger model is a particularly interesting system (see 
\cite{Wenger:2008tq,Wolff:2007ip,Chandrasekharan:2002ex,Gattringer:1999hr,Gattringer:1998cd,Gausterer:1995np,Scharnhorst:1995nf,Karsch:1994ae,Gausterer:1992np,Salmhofer:1991cc} for related examples)
since in the conventional representation 
it suffers from a complex action problem coming from two sources: a topological term, as well as finite density 
(in the two flavor model). Furthermore, the 
relativistic fermions lead to additional minus signs in the dualization. For the massless case with staggered fermions these problems 
were overcome in \cite{Gattringer:2015nea} and in this presentation we discuss possible Monte Carlo update algorithms and explore the 
challenges that arise when simulating fermion worldlines coupled to gauge degrees of freedom. We discuss the behavior of the theory in the 
continuum limit, in particular the emergence of the expected independence from the vacuum angle (in the massless case), as well as 
condensation phenomena with staggered fermions at finite chemical potential.


\section{The one-flavor model with topological term}
The partition function of the Schwinger model with topological term is given by 
\begin{equation}
Z=\int \!\!D[U] \int \!\!D[\bar{\psi}, \psi] \ e^{\, - \, S_G[U] \, - \, i \theta Q[U] \, -\, S_\psi[U,\, \bar{\psi},\, \psi]} \; .
\end{equation}
The degrees of freedom are $U(1)$-valued link variables $U_\nu (n)$ and one-component Grassmann valued staggered fermion 
variables $\psi(n)$ and $\bar{\psi}(n)$ sitting on the sites of a 2-dimensional $N_S \times N_T$ lattice. With $n=(n_1,n_2)$ we label the 
sites and with $\nu \in \{1,2 \}$ the Euclidean space $(\nu=1)$ and time $(\nu=2)$ directions. We use periodic boundary conditions for all
variables, except for the temporal boundary conditions of the fermions which are anti-periodic. In the partition sum $Z$ we integrate over 
the link variables with the product measure $D[U]$ of $U(1)$ Haar-measures and over the fermions with the Grassmann product measure 
$D[\bar{\psi}, \psi]$. The exponential of the Boltzmann factor consists of three terms. The staggered fermion action
\begin{equation}
S_\psi[U,\bar{\psi}, \psi] = \frac{1}{2} \sum_{n,\nu} \gamma_\nu(n)  
\left[ 	U_\nu(n) \bar{\psi}(n) \psi(n+\hat{\nu}) - U_\nu(n)^{-1} \bar{\psi}(n+\hat{\nu}) \psi(n) \right],
\end{equation}
where $\gamma_\nu(n) = (-1)^{\delta_{\nu,2} \ n_1}$ denotes the staggered sign function. The Wilson gauge action, 
\begin{equation}
S_G[U] = - \, \beta \sum_n \text{Re} \ U_p(n) \; , \quad 
U_p(n) = U_1(n) U_2(n+\hat{1}) U_1(n+\hat{2})^{-1} U_2(n)^{-1} \; ,
\end{equation}
with inverse gauge coupling $\beta$, and the topological charge
\begin{equation}
\label{equ:topo_charge}
Q[U] = \frac{1}{i 4\pi} \sum_n \left[ U_p(n) - U_p(n)^{-1} \right], 
\end{equation}
which is coupled via the vacuum angle $\theta$. The topological term introduces a complex action problem which 
prevents us from using Monte Carlo techniques in the conventional representation.

An elegant way to solve the complex action problem and to make non-vanishing $\theta$ accessible in a Monte Carlo simulation is the 
exact transformation of the partition sum to new degrees of freedom as shown in \cite{Gattringer:2015nea}. 
In this new representation the partition sum reads
\begin{equation}
\label{partition_sum_one_flavor}
	Z \; = \; \left( \frac{1}{2} \right)^V \sum_{ \left\{ l,d,p \right\} } \prod_n I_{|p(n)|} 
	\left(2 \sqrt{\eta	\bar{\eta}}\right) \left( \sqrt{\frac{\eta}{\bar{\eta}}} \ \right)^{p(n)}.
\end{equation}
The sum $\sum_{ \left\{ l,d,p \right\} }$ runs over all admissible configurations of the dual degrees of freedom, i.e., the loop occupation 
numbers $l(n,\nu) \in \{-1,0,1\}$ and dimer occupation numbers $d(n,\nu) \in \{0,1\}$ sitting on the links $(n,\nu)$ of the lattice, as well as 
the plaquette occupation numbers $p(n) \in [-\infty, \infty]$ attached to the bottom left lattice site $n$ of a plaquette. 
$V = N_S N_T$ denotes the total number of lattice sites, which we refer to as the volume of the lattice. In the dual form all variables have periodic 
boundary conditions in both directions. $I_n (x)$ denotes the modified Bessel functions of the first kind, and we 
introduced the abbreviations $\eta = \frac{\beta}{2}-\frac{\theta}{4 \pi}$ and $\bar{\eta} = \frac{\beta}{2}+\frac{\theta}{4 \pi}$.

\begin{figure}[t]
	\centering
	\begin{subfigure}{0.5\textwidth}
		\raggedright
		\includegraphics[width=0.95\textwidth]{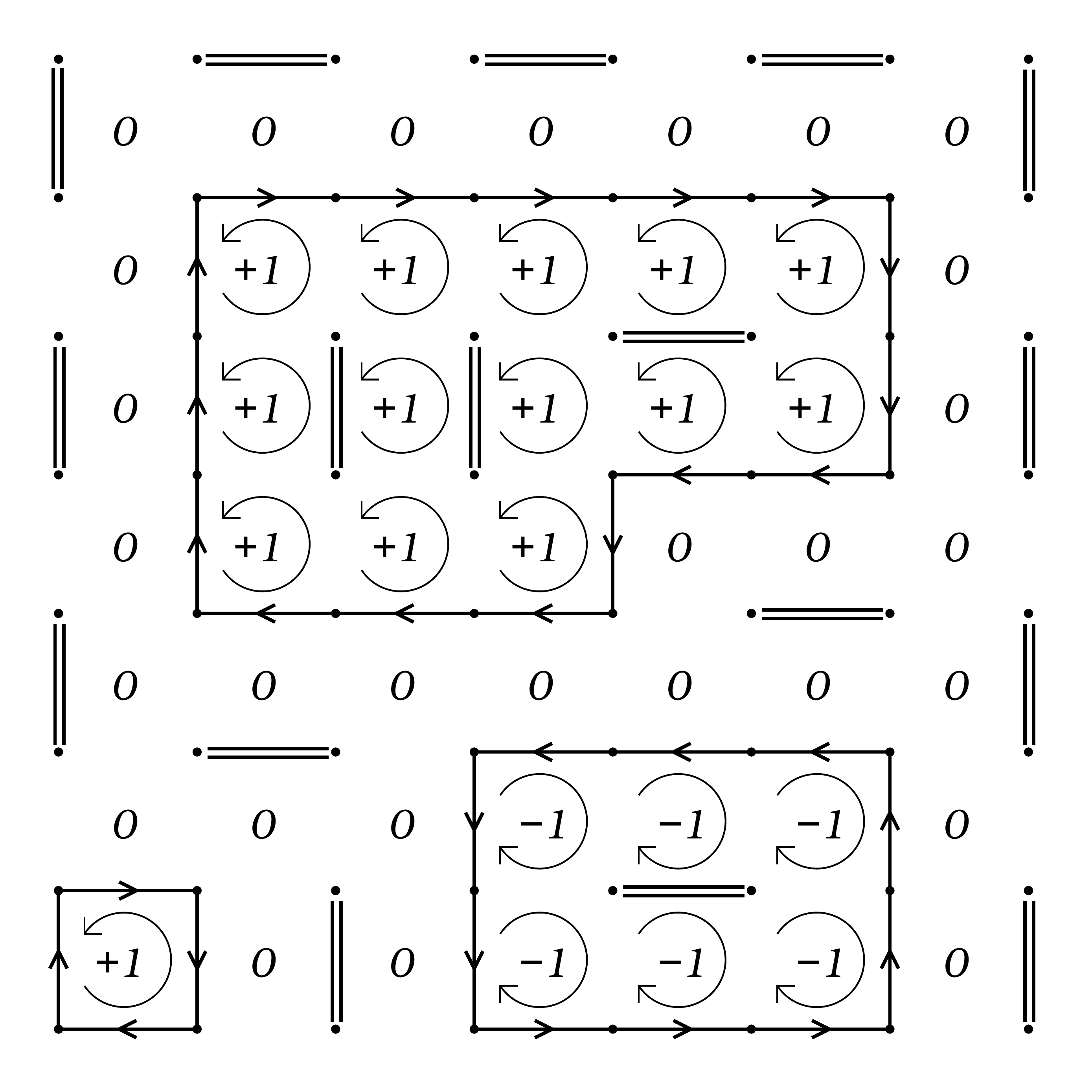}
	\end{subfigure}%
	\begin{subfigure}{0.5\textwidth}
		\raggedleft
		\includegraphics[width=0.95\textwidth]{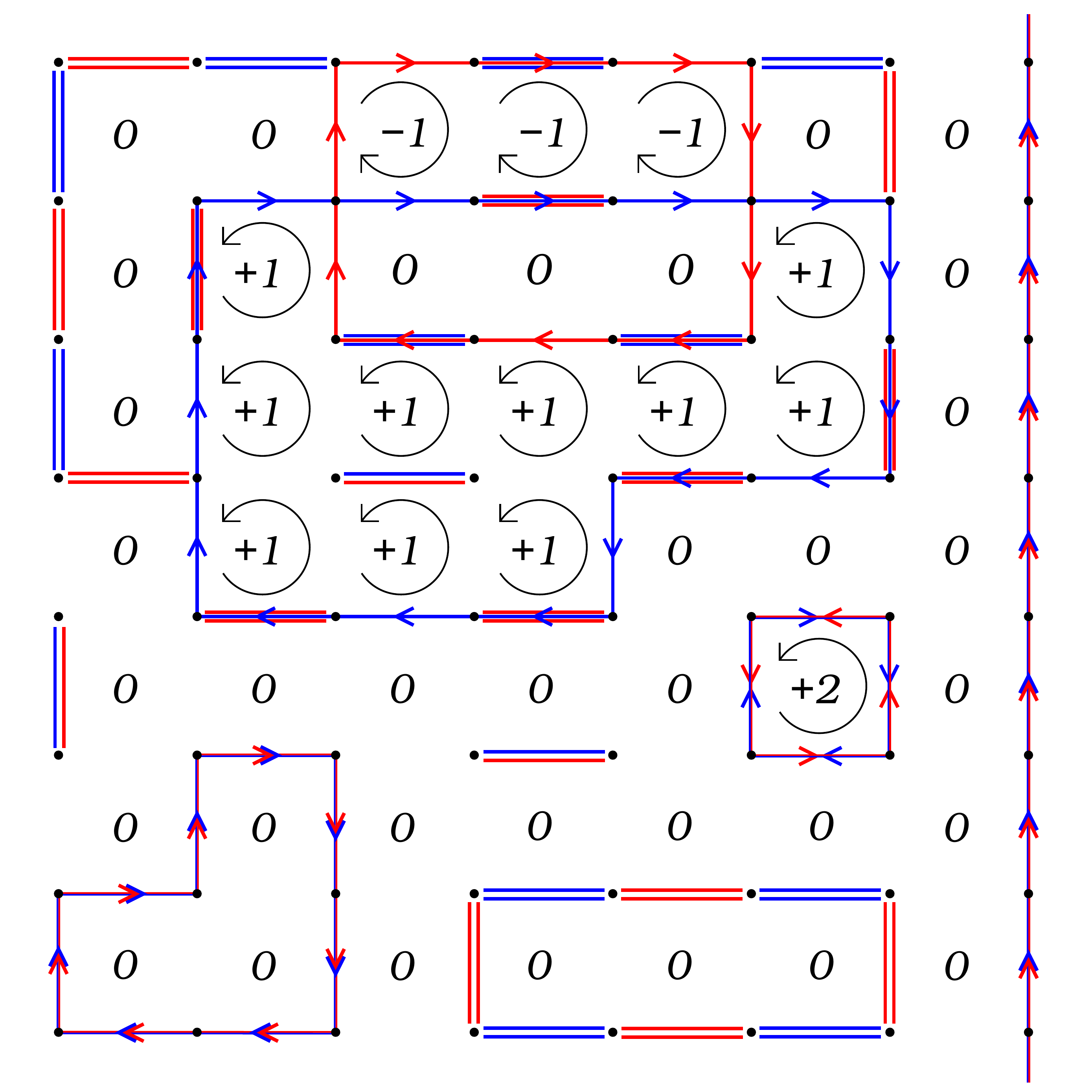}
	\end{subfigure}%
	\caption{Examples of admissible configurations on a $8\times 8$ lattice for the one-flavor model (lhs.) and the two-flavor model 
	(rhs.), where red and blue distinguishes the two flavors. Double lines denote dimers, single lines denote loops and the circular 
	arrows represent plaquettes with the plaquette occupation numbers written inside explicitly (plots from 
	\cite{Gattringer:2015nea,Goschl:2017kml}).}
	\label{admissible_configuration}
\end{figure}

The configurations of the dual variables are subject to two constraints: 1) The Pauli principle demands that each lattice point has to be 
either the endpoint of a dimer or to be run through by a loop. 2) At each link the loop flux and the flux from occupied plaquettes 
has to sum up to zero. An example for an admissible configuration in the one-flavor model is shown in the lhs.\ plot of 
Fig.~\ref{admissible_configuration}. Double lines represent dimers and single lines with arrows correspond to the loops. The arrows
on the loops indicate the orientation of the loops corresponding to the values $\pm 1$ of the loop occupation numbers $l(n,\nu)$. 
An arrow pointing in positive direction, i.e., upwards or to the right, corresponds to a link occupation number of $+1$, while an arrow 
pointing in negative direction corresponds to a link occupation number of $-1$. The plaquette occupation numbers are explicitly shown 
inside the plaquettes with a circular arrow indicating the orientation of the plaquette. 

In the simulation we will compute bulk observables which can be obtained as the first and second derivatives of $\ln(Z)$. In particular we 
focus on the plaquette expectation value $\langle U_p \rangle$ and its susceptibility $\chi_p$, as well as the topological charge density $
\langle q \rangle$ and the topological susceptibility $\chi_t$,
\begin{equation}
\langle U_p \rangle \; = \; \frac{1}{V} \frac{\partial}{\partial \beta} \ln(Z) \; \; ,  \; \; 
\chi_p \; = \; \frac{1}{V} \frac{\partial^2}{\partial\beta^2} \ln(Z) \; \; ,  \; \; 
\langle q \rangle  \; = \; \frac{1}{V} \frac{\partial}{\partial \theta} \ln(Z) \; \; ,  \; \; 
\chi_t  \; = \; \frac{1}{V} \frac{\partial^2}{\partial\theta^2} \ln(Z) \; .
\end{equation}

\section{Two-flavor model with chemical potential}
In the two-flavor model with chemical potential the partition sum is given by
\begin{equation}
\label{partition_sum_2flavor_conventional}
Z \; = \; \int \! D[U] \, D[\bar{\psi}, \psi] \, D[\bar{\chi}, \chi] \ e^{\, -\, S_G[U] \, -i \, \theta Q[U] 
\, - \, S_\psi [U,\, \bar{\psi},\, \psi] \, - \, S_\chi [U,\, \bar{\chi},\, \chi]} \;.
\end{equation}
The variables $\psi$ and $\chi$ represent positively and negatively charged fields respectively. The measure $D[\bar{\chi},\chi]$ 
for the second flavor is again a product over Grassmann integrals. The fermionic actions for the two flavors contain the chemical potentials $\mu_\phi$ and $\mu_\chi$:
\begin{align}
S_\psi [U,\bar{\psi},\psi] = &\frac{1}{2} \sum_{n,\nu} \gamma_\nu(n) 
\left[  \ e^{\, \mu_\psi \delta_{\nu,2}} \, U_\nu (n) \, \bar{\psi}(n) \, \psi (n+\hat{\nu}) -  
\ e^{\, -\mu_\psi \delta_{\nu,2}} \ U_\nu (n)^{-1} \, \bar{\psi}(n+\hat{\nu}) \, \psi (n)    \right], \\
S_\chi [U,\bar{\chi},\chi] = &\frac{1}{2} \sum_{n,\nu} \gamma_\nu(n) 
\left[  \ e^{\, \mu_\chi \delta_{\nu,2}} \ U_\nu (n)^{-1} \, \bar{\chi}(n) \, \chi (n+\hat{\nu}) -  
\ e^{\, -\mu_\chi \delta_{\nu,2}} \, U_\nu (n) \, \bar{\chi}(n+\hat{\nu}) \, \chi (n)    \right].
\end{align}
The crucial difference of the two fermion actions is that we use the complex conjugate link variables $U_\nu (n)^\ast = U_\nu (n)^{-1}$ for 
the second flavor $\chi$, implying opposite charges for $\psi$ and $\chi$. This implements Gauss' law, which requires overall electric 
neutrality. Non-zero values of the chemical potentials also generate a complex action problem.

Using dualization \cite{Gattringer:2015nea} we obtain the flux representation of the two flavor partition sum, 
\begin{equation}
\label{partition_sum_2flavor_dual}
Z \, =\, \left( \frac{1}{2} \right)^{2 V} \sum_{\{l,d,\bar{l},\bar{d},p\}} e^{\, \mu_\psi N_T W(l)} \ e^{\, \mu_\chi N_T W(\bar{l})}
	\prod_n I_{|p(n)|} \left(2 \sqrt{\eta \bar{\eta}}\right) \left( \! \sqrt{\frac{\eta}{\bar{\eta}}} \right)^{p(n)}.
\end{equation}
For the second flavor $\chi$ we need a second loop variable $\bar{l}$ and a second dimer variable $\bar{d}$. The sum runs over all 
admissible configurations of the loop ($l, \bar{l}$) and dimer ($d, \bar{d}$) occupation numbers of both flavors and the plaquette 
occupation number $p$. For both flavors each site of the lattice has to be either the endpoint of a dimer or part of a loop. The flavors 
interact via the plaquette occupation numbers $p$, which now have to compensate the combined link flux of both flavors. The flux from 
the loops $\bar{l}$ is counted with a negative sign as it represents the negatively charged field $\chi$. $W(l)$ and $W(\bar{l})$ are the 
temporal net 
winding numbers of the loops of the two flavors. The respective exponential factors $e^{\, \mu_\psi N_T W(l)}$ ($e^{\, \mu_\chi N_T 
W(\bar{l})}$) only contribute for loops winding around the compact time. In the one-flavor case, this factor vanishes automatically 
since we always have an equal number of forward and backward winding loops and, thus, a net winding number of zero. In the two-flavor 
case, the second flavor allows non-vanishing net winding numbers, however, $W(l)=W(\bar{l})$ as required by Gauss' law. 
In the rhs.\ plot of  
Fig.~\ref{admissible_configuration} we show an example of an admissible configuration on a $8 \times 8$ lattice. The two flavors are 
distinguished by using blue and red color. Along the right edge we see a simple 
combination of two winding loops where loops of both flavors sit on top of each other. We refer to these loops as ''neutral double loops'',
and stress that the same orientation of the loops is necessary to meet the constraint of a vanishing link 
flux. The temporal winding numbers for our example configuration are $W(l) = W(\bar{l}) = +1$.  
The chemical potentials give rise to additional bulk 
observables, i.e., the particle number densities $n_{\psi/\chi}$ and the corresponding susceptibilities $\chi_{n_{\psi/\chi}}$,
\begin{equation}
n_{\psi/\chi} = \frac{1}{V} \frac{\partial \text{ln}(Z)}{\partial (\mu_{\psi/\chi} N_T)} \ , \qquad
\chi_{n_{\psi/\chi}} = \frac{1}{V} \frac{\partial^2 \text{ln}(Z)}{\partial (\mu_{\psi/\chi} N_T)^2} \ .
\end{equation}

\section{Update strategies}
Since all weights are real and positive in the dual representations (\ref{partition_sum_one_flavor}) and 
(\ref{partition_sum_2flavor_dual}) the dual formulation solves the complex action problem. However, the dual partition sums 
constitute highly constrained systems and for ergodicity we need different types of updates which also respect the constraints. 
For the one-flavor case we use local updates for the loops and plaquette variables and a worm update for the dimers. 
In the two-flavor model we use the updates of the one-flavor system independently for both flavors, 
but also need an additional worm update for the winding neutral double loops. For faster propagation through phase space we also use 
global gauge/plaquette updates for both cases and local updates for electrically neutral loops in the two-flavor case. 

\begin{figure}[t]
\centering
\includegraphics[width=\textwidth]{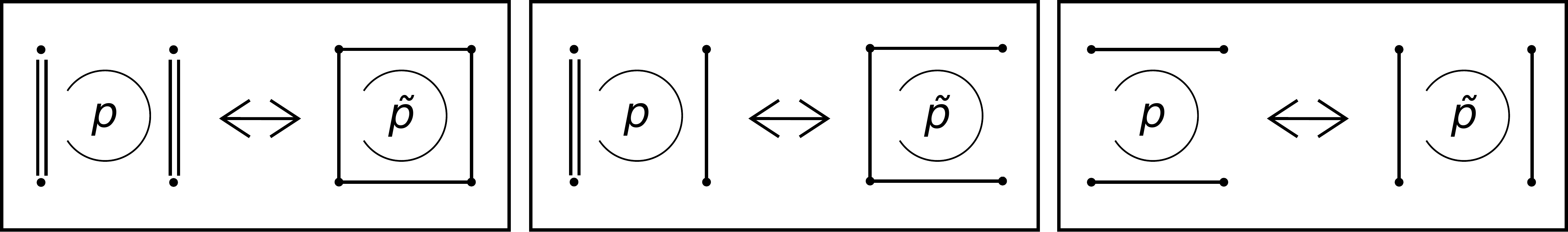}
\caption{Schematic illustration of the local updates. In the one-flavor model, as well as for the single flavor updates of the 
two-flavor model, the double lines denote dimers and the single lines denote fermion loops. In the two-flavor case 
the same steps are also used for double dimers and identical loops of both flavors sitting on top of each other (neutral double loops). 
The orientations of the loops are not shown here, but is is trivial to add them and to determine the necessary change of the 
plaquette occupation numbers from $p$ to $\tilde{p}$ such that changed flux from fermion loops is compensated by changing the 
plaquette occupation number. The three types of local updates are:  Changing between a pair of dimers and a one-plaquette loop 
(lhs.\ plot), growing/shrinking a loop by absorbing/adding a dimer (center), or joining/separating two loops (rhs.).}
\label{local_update_scheme}
\end{figure}

\subsection{Local updates}

The local updates change the gauge and fermionic degrees of freedom on a single plaquette. There are three possibilities which we 
illustrate in Fig.~\ref{local_update_scheme}. The local updates change dimers and loops, where the loops can either be charged 
loops of a single flavor or (only in the two-flavor model) neutral double loops where both flavors propagate together. Accordingly, the 
dimers in Fig.~\ref{local_update_scheme} are either dimers of one flavor or dimers of both flavors sitting on top of each other.
The orientation of the loops is not shown to keep the illustration simple, but it is easy to add it for the charged loops and to determine the 
corresponding change of the plaquette occupation numbers from $p$ to $\tilde{p}$ such that also the new configuration has vanishing
flux at each link.  

The first local update (lhs.\ plot in Fig.~\ref{local_update_scheme}) is to switch between a pair of dimers and a 
loop around a plaquette with random orientation. The second local update (center plot) expands an existing loop by 
absorbing a dimer sitting parallel to a loop segment or shrinks a loop by placing a dimer. The third option (rhs.\ plot) 
either combines two neighboring loops with the same orientation or separates a loop into two loops with the same orientation.

Since neutral double loops do not generate any link flux, local neutral double loop updates do not change the plaquette occupation 
numbers and, therefore, are always accepted. Local updates for charged loops require a change of the plaquette occupation 
numbers to compensate the additional flux and we accept the update in a Metropolis step with acceptance 
probability min$\{\rho_\pm, 1\}$ where
\begin{equation}
\rho_+ = \sqrt{\frac{\eta}{\bar{\eta}}} \ \frac{I_{|p(n)+1|} (2 \sqrt{\eta \bar{\eta}})}{I_{|p(n)|} (2 \sqrt{\eta \bar{\eta}})} \ , \qquad
\rho_- = \sqrt{\frac{\bar{\eta}}{\eta}} \ \frac{I_{|p(n)-1|} (2 \sqrt{\eta \bar{\eta}})}{I_{|p(n)|} (2 \sqrt{\eta \bar{\eta}})} \ .
\end{equation}

\subsection{Worm update for the dimers}

The local updates discussed above are not ergodic: As discussed in \cite{KASTELEYN19611209} for pure dimer configurations on a 
two-dimensional lattice with periodic boundary conditions (which is an admissible configuration of our models) the dimer configurations 
come in four different topological sectors which are not connected by local moves and a non-local update is needed. The non-local 
update of choice is a worm that searches for closed non self-intersecting loops of links such that along the loop empty links
alternate with links occupied by dimers. Once the worm reaches its starting point, all dimers along the loop are shifted by one link.  
This move modifies the dimers only, leaving the loops and plaquettes untouched. Thus, also the weight of the configuration stays invariant 
and no Metropolis decision is needed for accepting a dimer modification. 

\subsection{Updates for winding neutral double loops}
\label{sec:update_winding}
The two-flavor case also includes configurations of neutral double loops with non-vanishing net winding numbers, 
which are not accessible with local loop updates. Therefore we use an additional worm-based strategy which identifies 
closed chains of alternating dimers of both colors (see, e.g., the bottom of the rhs.\ plot of Fig.~\ref{admissible_configuration}) 
and switches between those and neutral double loops. The plaquette occupation numbers do not change for such a step 
and a Metropolis decision is only necessary if the neutral double loop winds around compact time. The corresponding 
Metropolis acceptance probability then is min$\{ 1,\exp(W(\mu_\psi+\mu_\chi)N_T) \}$, where $W$ is the proposed change 
of the temporal winding number. 

\section{Results and discussion}

\begin{figure}[t]
\centering
\includegraphics[width=\textwidth]{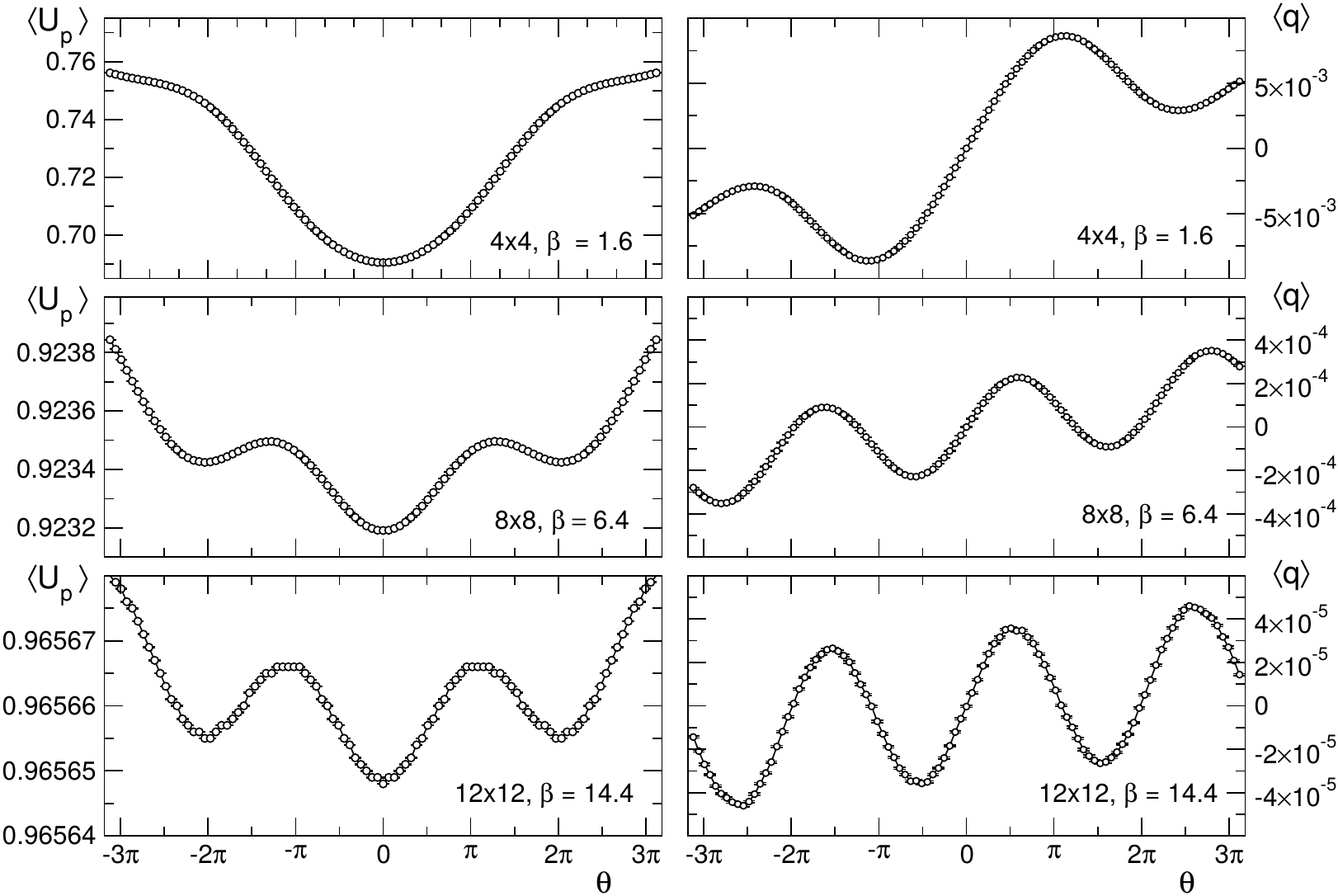}
\caption{The plaquette expectation value (lhs.) and the topological charge density (rhs.) as a function of $\theta$ (figures from
\cite{Goschl:2017kml}). From top to bottom we approach the continuum limit by increasing the lattice volume and $\beta$ 
while keeping the ratio $R=\beta/N_S N_T$ constant at $R=0.1$.}
\label{theta_dependence}
\end{figure}

\subsection{Results for the one-flavor model}

The simulation of the one-flavor model was carefully verified against a standard simulation for vanishing vacuum angle $\theta$ and 
against an exact summation on small lattices. The main physical question was to establish the emergence of the correct 
$\theta$-dependence of the observables for the formulation with staggered fermions and the field theoretical definition 
of the topological charge as defined in (\ref{equ:topo_charge}). In scalar QED$_2$ one observes a $2\pi$ periodicity in $\theta$ when 
approaching the continuum limit \cite{Gattringer:2015baa}. 
This behavior is also manifest in the Schwinger model as shown in Fig.~\ref{theta_dependence}. We plot 
the topological charge density (lhs.) and the plaquette expectation value (rhs.) as a function of $\theta$. From top to bottom we 
approach the continuum limit by increasing the lattice volume and $\beta$ while keeping the ratio $\beta / N_S N_T = 0.1$ fixed.
We observe that both observables indeed recover the $2\pi$-periodicity as expected for the continuum limit of the field theoretical 
definition of $Q[U]$. 

A second important aspect concerns the amplitude of the $\theta$-dependent oscillations seen in Fig.~\ref{theta_dependence}. 
The $2D$ lattice Schwinger model with staggered fermions has to be compared to the two-flavor continuum model. There, the 
observables will be independent of $\theta$ if one of the fermion masses vanishes as we can always use a chiral rotation and the 
anomaly of the fermion determinant to remove the vacuum angle. This behavior can also be observed in Fig.~\ref{theta_dependence},
where $\langle U_P \rangle$ resembles a parabola and $\langle q \rangle$ an inclined straight line, both overlaid with oscillations. As we 
approach the continuum limit, the amplitude of the oscillations as well as the curvature of the underlying parabola in 
$\langle U_P \rangle$ and the overall slope in $\langle q \rangle$ decrease quickly. This suggests that the $\theta$-dependence vanishes 
in the continuum limit and that staggered fermions implement the anomaly correctly (compare also \cite{Durr:2004ta,Durr:2003xs}).

\subsection{Results for the two-flavor model}

\begin{figure}[t]
	\centering
	\begin{subfigure}{0.59\textwidth}
		\centering
		\includegraphics[width=\textwidth]{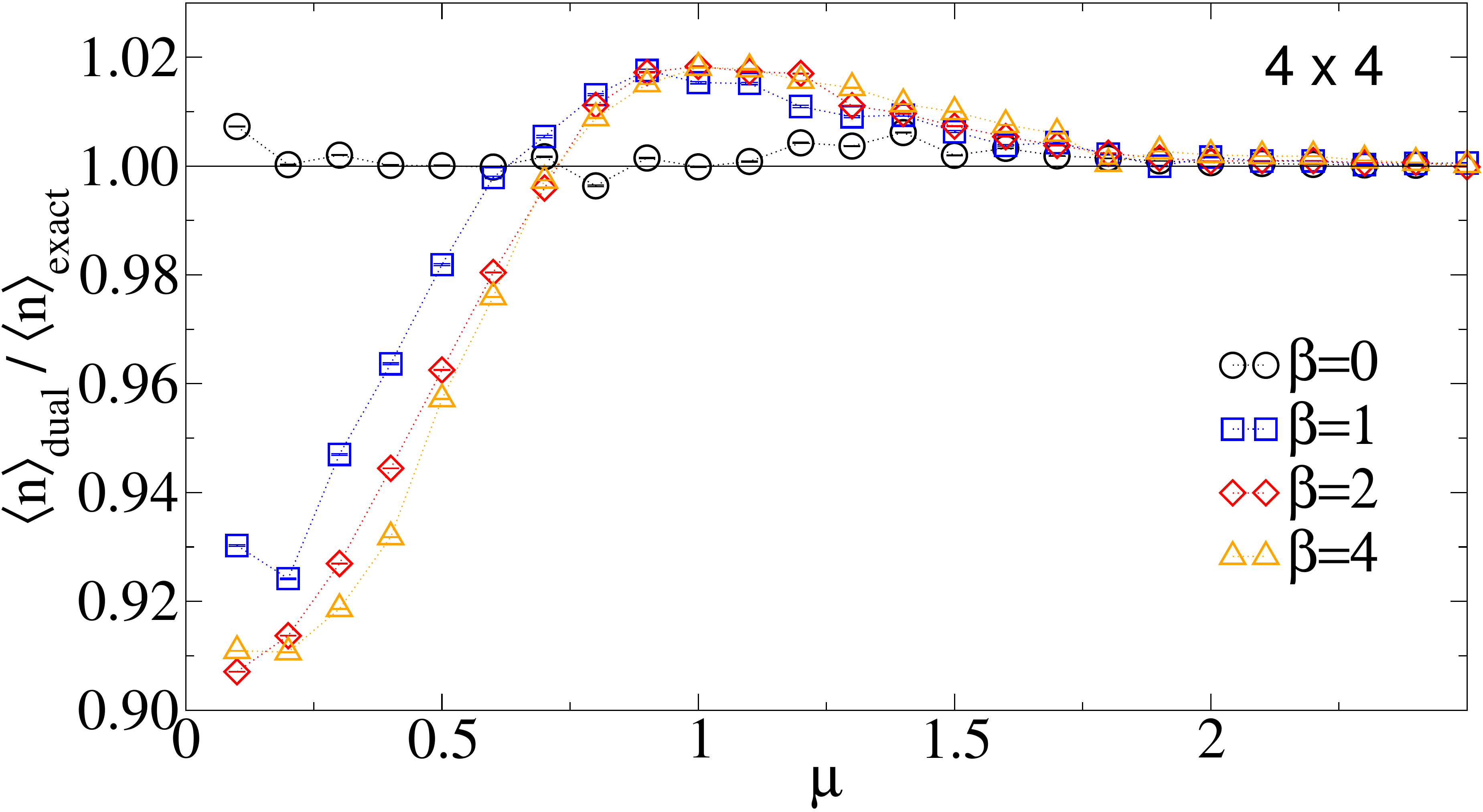}
	\end{subfigure}%
	\hspace{10mm}
	\begin{subfigure}{0.31\textwidth}
		\centering
		\includegraphics[width=\textwidth]{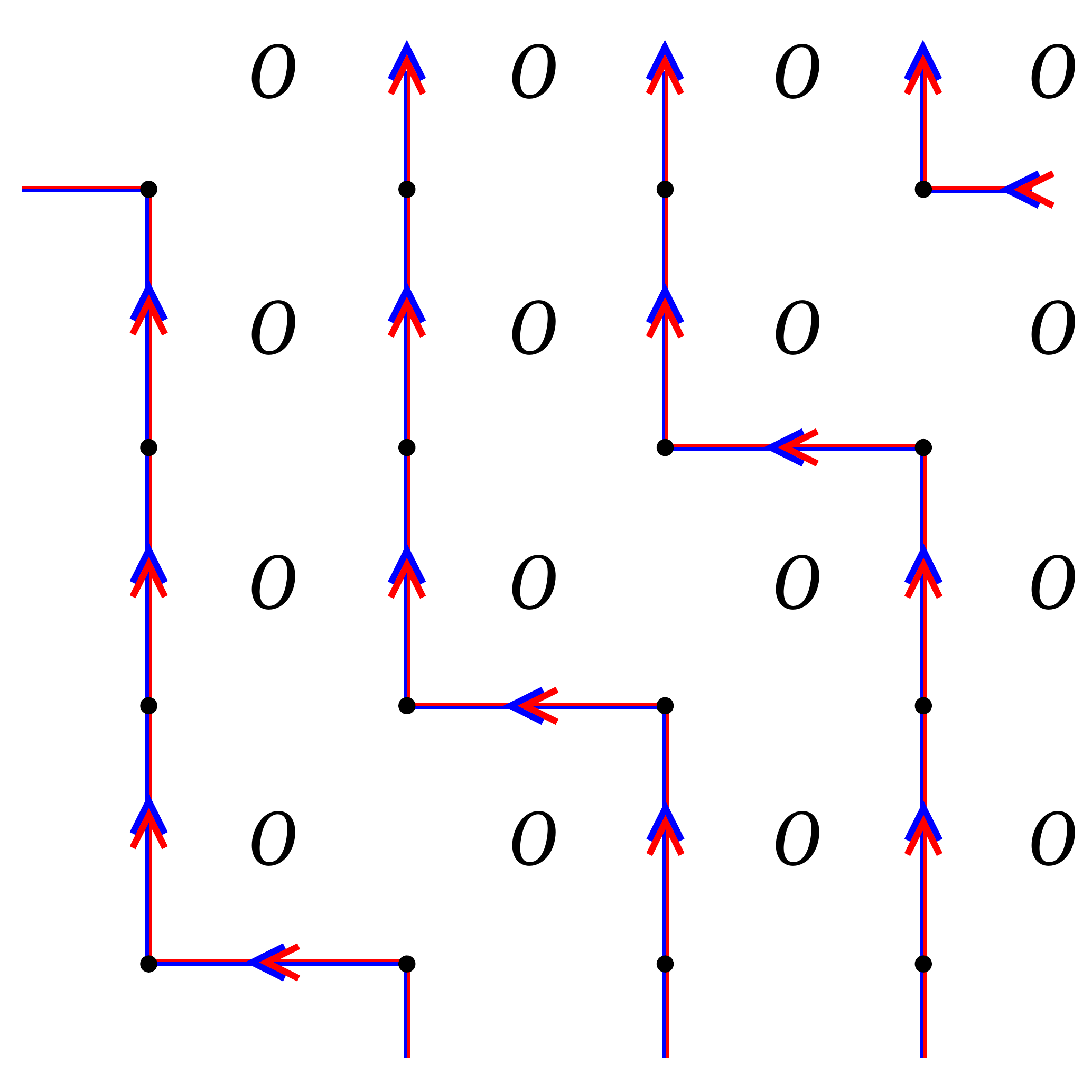}
	\end{subfigure}%
\caption{Lhs.: The ratio of particle number densities $\langle n \rangle_{dual}/ \langle n \rangle_{exact}$ versus $\mu$, for different values 
of $\beta$. The solid black line shows the correct ratio of 1 and the dotted lines connect the data points to guide the eye. 
Rhs.: Example of a topologically stabilized configuration of a neutral double loop. It winds once around the spatial and 
three times around the temporal boundary, filling the entire lattice.}
\label{autocorrelation}
\end{figure}

In the two-flavor model with chemical potentials we can access particle number densities as additional observables. 
For a small lattice, we can again employ an exact summation to test the dual simulation. 
This comparison is shown in the lhs.\ of Fig.~\ref{autocorrelation}, where we plot the ratio of the dual results and 
the exact data on a $4\times 4$ lattice as a function of $\mu$ and for different values of $\beta$. We see that for vanishing 
$\beta$ we get acceptable results, however, with increasing $\beta$ a systematic discrepancy emerges
for $\mu<2$. The reason is that with increasing $\beta$ it is more likely that activated plaquettes surrounded by charged
loops appear, which then make it more difficult to insert winding neutral double loops that couple to the chemical potential.
This leads to longer autocorrelations for $\mu$-dependent quantities. At $\mu>2$ the particle numbers start to saturate at 
their maximum, admitting neither large fluctuations of the winding number nor strong local variations of plaquettes and the 
effect of short local loops is reduced.

However, for large $\mu$ (and thus high winding numbers) a second effect can be observed that leads to long autocorrelations: 
For high winding numbers topologically stabilized configurations appear which are a considerable obstacle for the two-flavor 
simulation and lead to large autocorrelation times. An example for such a configuration is given in the rhs.\ plot of 
Fig.~\ref{autocorrelation} where we see the example of a single neutral double loop that covers the entire lattice resulting 
in a winding number of $+3$. At large $\mu$ the increase of the winding number to its maximum is strongly favored, 
yet topologically not possible. With the given update strategies the only possibility to alter this configuration is to delete 
the double loop. This, however, is strongly suppressed by the Boltzmann factor as it comes with a decrease of the winding 
number by a value of $3$. Similar topologically stabilized configurations can be found at arbitrary values of $\beta$.
For larger lattices, the combination of these two problems (local loops blocking winding loops and topologically 
stabilized configurations) lead to large autocorrelations and enormous computational effort. However, we expect the 
problems to be much milder for lattices of dimension $D\geq 3$, because there, (the one-dimensional) winding loops 
have more dimensions to propagate and are less likely to block each other.

\section{Conclusion}
In this presentation we explore the dual formulation of the massless lattice Schwinger model for its use in Monte Carlo simulations. 
Although the dual representation has only real and positive weights and the complex action problem is solved,
we have to deal with highly constrained systems of fermion loops and dimers. Several different update steps are necessary to 
guarantee ergodicity and we verify our algorithm against conventional simulations and exact summation on small lattices. 
For the one-flavor model we study the approach to the correct continuum limit and establish the $\theta$-independence 
of the observables, as expected in the massless case. In the two-flavor model we have to augment the algorithms by updating
neutral double loops winding around time with a worm strategy. We identify two sources of problems for a Monte Carlo simulation,
which both can be attributed to the Pauli principle. The fact that loops of the same flavor are not allowed to intersect gives rise to a 
very stiff system in two dimensions. At larger values of $\mu$ we see the emergence of topologically stabilized configurations 
which are hard to update and lead to very long autocorrelations. For the particularly challenging two-dimensional case
the recently proposed canonical worldline approach \cite{Orasch:2017niz} could be a powerful alternative which we currently explore.

\vskip3mm
\noindent
\textbf{Acknowledgments:} The author thanks Christof Gattringer, Alexander Lehmann and Christoph Weis for discussions and 
support, as well as for providing parts of the reference data on small volumes. He is supported by the FWF DK W 1203, "Hadrons in Vacuum, Nuclei and Stars", and by the FWF project I 2886-N27 in cooperation with DFG TR55, ''Hadron Properties from Lattice QCD''. 
\bibliography{lattice2017}

\end{document}